\documentclass[letterpaper]{article}
\usepackage{aaai2027}
\usepackage[hyphens]{url}
\usepackage{graphicx}
\usepackage{natbib}
\usepackage{caption}
\usepackage{algorithm}
\usepackage{algorithmic}
\usepackage{booktabs}
\usepackage{amsmath}
\usepackage{amssymb}
\usepackage{multirow}

\nocopyright

\newcommand{\benchmarkname}{\textsc{PonyEval}}

\title{\benchmarkname{}: Evaluating LLM-Based Program Repair for Capability-Safe and Actor-Oriented Pony Software}
\author{
Bang Xie\textsuperscript{\rm 1}, \quad Hao Liu\textsuperscript{\rm 1}, \quad
Zhenyu Shi\textsuperscript{\rm 1}, \quad Zhiyuan Peng\textsuperscript{\rm 1},\\
Xin Yin\textsuperscript{\rm 2}, \quad Chenhao Ying\textsuperscript{\rm 1}\corresponding, \quad
Yuan Luo\textsuperscript{\rm 1}, \quad Haiming Jin\textsuperscript{\rm 1},\\
Wei Chen\textsuperscript{\rm 1}, \quad Senjian Zhang\textsuperscript{\rm 1}, \quad
Shaocong Long\textsuperscript{\rm 1}
}
\affiliations{
\textsuperscript{\rm 1}Shanghai Jiao Tong University, Shanghai, China\\
\textsuperscript{\rm 2}Zhejiang University, Hangzhou, China\\
yingchenhao@sjtu.edu.cn
}

\begin{document}
\maketitle

\begin{abstract}
Repository-level issue-resolution benchmarks have made executable evaluation central to software-engineering agents, but their language coverage remains concentrated in mainstream ecosystems.
Pony presents a different regime: it combines actors, reference capabilities, ahead-of-time compilation, and a rapidly evolving historical toolchain, making both patch generation and faithful replay difficult.
We introduce \benchmarkname{}, a SWE-bench-style benchmark of 291 real GitHub issue--pull-request pairs from 15 Pony repositories.
Every instance binds an issue statement, a historical base commit, a developer gold patch, a black-box test patch, and a reproducible runtime mapping.
The frozen release passes an offline audit requiring the issue-specific test to fail on the base state and pass after the gold patch; it contains no duplicate instance identifiers or canonical repository--PR pairs.
In a separate full-set semantic selection audit, three isolated machine reviewers label all 291 instances as include or exclude; their Fleiss' $\kappa$ is 0.8968, with 249 unanimous inclusions and 32 unanimous exclusions.
To replay eleven years of repository history, we reconstruct 72 runtime images covering 289 unique base commits and verify their availability on five heterogeneous compute nodes.
We define a matched evaluation with mini-SWE-agent 2.4.6 for GPT-5.6-sol, DeepSeek-V4-Pro, GLM-5.2, MiniMax-M3, and Kimi-K3, followed by strict patch application, compilation, and hidden-test validation.
Across the patches actually produced by each model, conditional resolution rates range from 10.21\% to 24.68\%.
These rates characterize the quality of generated patches rather than success over all 291 benchmark tasks.
\end{abstract}

\section{Introduction}

Real software repair requires more than synthesizing a short function.
An agent must interpret a natural-language issue, navigate a repository, identify the relevant implementation, edit code under project-specific conventions, and produce a patch that survives execution.
SWE-bench operationalized this setting through historical GitHub issues and corresponding pull requests \cite{jimenez2024swebench}; subsequent systems showed that an interactive agent--computer interface can substantially change repair performance \cite{yang2024sweagent}.
Yet the resulting empirical picture is still shaped by languages and repositories for which installation, testing, and training data are abundant.

Recent multilingual benchmarks broaden issue resolution beyond Python \cite{zan2025multiswebench}, but long-tail language ecosystems remain underrepresented.
This omission matters because a benchmark is not merely a collection of source files.
It requires a version-correct compiler, dependencies that still resolve, an issue-specific oracle, and a validator that distinguishes semantic failure from build, provider, and infrastructure failure.
Those requirements are especially demanding when a language's type system and runtime encode behavior that a model cannot approximate with generic syntax.

Pony is an object-oriented actor language designed around reference capabilities and static data-race freedom \cite{clebsch2015deny}.
Its runtime and type system were co-designed for actor-local garbage collection and message tracing \cite{clebsch2017orca}.
These properties make Pony an informative target for repository-level repair: relevant defects span capability typing, compiler internals, actor scheduling, foreign-function interfaces, libraries, and historical build systems.
They also make naive replay unreliable.
A patch that looks plausible may fail under the exact compiler expected by the historical commit, while a test executed under a modern compiler may no longer measure the original defect.

We present \benchmarkname{}, an executable benchmark built from real Pony GitHub issues and their fixing pull requests.
Each instance packages the natural-language task with a base commit, gold patch, black-box test patch, and provenance needed to reproduce the historical execution.
Following the fail-to-pass principle used for issue reproduction in SWT-Bench \cite{mundler2024swtbench}, an instance is admitted only when the same issue-specific test fails on the base state and passes after the developer patch.
When existing tests are missing or insufficiently isolated, the construction pipeline follows ArkEval's central idea of generating and independently reviewing executable oracles \cite{xie2026arkeval}; generated tests remain a small, explicitly marked minority of the final set.

The paper makes three contributions:
\begin{itemize}
    \item We introduce a 291-instance Pony issue-resolution benchmark covering 15 repositories and commits from 2015 to 2026, with gold patches, black-box test patches, and zero duplicates under both instance-ID and canonical repository--PR audits.
    \item We reconstruct a versioned execution layer with 72 historical runtime images for 289 unique base commits and verify all images on five heterogeneous nodes, including an explicit compatibility adapter for legacy glibc images.
    \item We report a matched five-model mini-SWE-agent evaluation of patch-conditional localization, application, compilation, and semantic resolution.
\end{itemize}

\begin{figure*}[t]
\centering
\includegraphics[width=\textwidth]{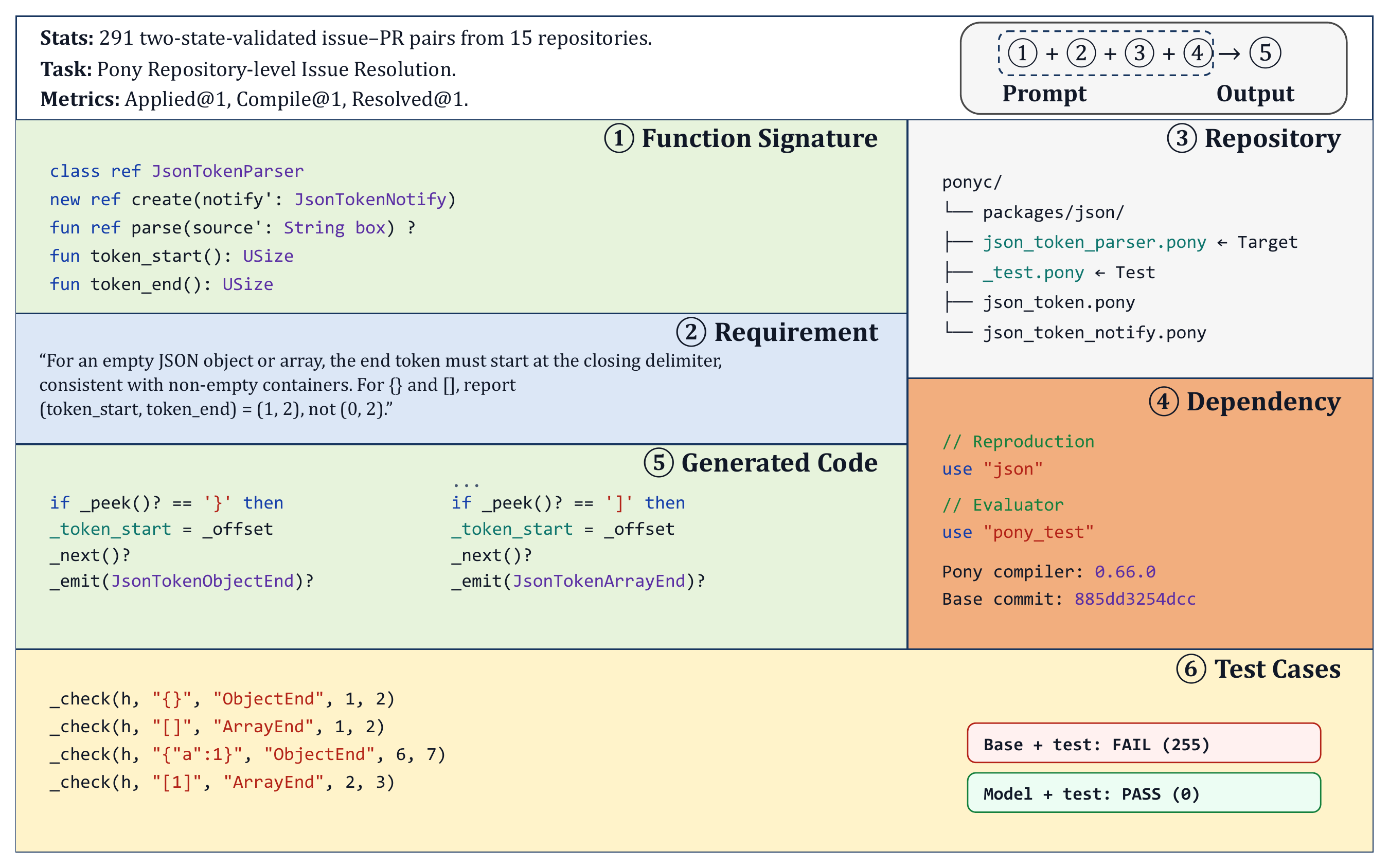}
\caption{Overview of the \benchmarkname{} task interface and executable evaluation workflow.}
\label{fig:overview}
\end{figure*}

\section{Background and Related Work}

\paragraph{Executable issue-resolution benchmarks.}
Defects4J and QuixBugs provide executable bug corpora widely used in repair evaluation \cite{just2014defects4j,lin2017quixbugs}.
Code-generation benchmarks cover standalone functions, competition problems, and data-science libraries \cite{chen2021evaluating,hendrycks2021measuring,lai2023ds1000}.
ClassEval evaluates interdependent methods within a class \cite{du2023classeval}; CoderEval and DevEval incorporate progressively broader file and repository context \cite{yu2024codereval,li2024deveval}.
CrossCodeEval and RepoBench isolate cross-file retrieval and completion within multilingual or repository-level settings \cite{ding2023crosscodeeval,liu2024repobench}.
SWE-bench frames repair as producing a repository patch for a real GitHub issue and evaluates it against executable tests \cite{jimenez2024swebench}.
SWE-agent demonstrates that the interface through which a model searches, edits, and executes a repository is itself an important experimental variable \cite{yang2024sweagent}.
Multi-SWE-bench extends issue resolution to seven non-Python languages and documents substantial cross-language variation \cite{zan2025multiswebench}.
Repository-generation benchmarks study adjacent but distinct constraints: SolEval evaluates repository-level Solidity generation, whereas RepoGenesis evaluates complete microservice repositories with deployment and cross-file consistency checks \cite{peng2025soleval,peng2026repogenesis}.
\benchmarkname{} complements these efforts by evaluating issue-driven repair rather than repository generation, studying a substantially smaller language ecosystem, and treating historical compiler reconstruction as part of the benchmark rather than as an incidental deployment detail.

\paragraph{Automated program repair.}
Automatic repair spans search-based methods such as GenProg \cite{weimer2009automatically,le2012genprog}, probabilistic models learned from correct code \cite{long2016automatic}, symbolic synthesis \cite{mechtaev2016angelix}, and learned template ranking \cite{saha2017elixir}; Monperrus surveys this broader design space \cite{monperrus2018automatic}.
Neural repair systems learn code transformations from historical fixes, with architectures that add context and code-aware constraints \cite{tufano2019empirical,lutellier2020coconut,jiang2021cure,li2020dlfix}.
BugLab instead trains a repair detector with bugs generated by a co-trained selector \cite{allamanis2021selfsupervised}.
For pretrained models, prior work evaluates Codex on QuixBugs \cite{prenner2021automatic}, conversational repair with ChatGPT \cite{xia2024conversation}, automated repair of LLM-generated programs \cite{fan2023automated}, and retrieval-augmented LLM repair \cite{jin2023inferfix}.
Unlike methods designed to maximize fixes on established Java or Python suites, \benchmarkname{} fixes the agent scaffold and studies whether candidate patches survive historical Pony compilation and issue-specific behavioral tests.

\paragraph{Issue-specific tests as oracles.}
SWT-Bench formalizes issue reproduction as a fail-to-pass test: a candidate test must fail before the gold fix and pass afterward \cite{mundler2024swtbench}.
ArkEval adapts this idea to ArkTS repositories that often lack adequate regression tests, using multiple agents to synthesize and review execution-based oracles \cite{xie2026arkeval}.
\benchmarkname{} adopts the same executable criterion, while prioritizing developer-written test changes when available and marking generated tests by origin.
The benchmark audit checks outcomes, provenance, and duplicate identities; it does not infer correctness from textual similarity to the gold patch.

\paragraph{Tool-augmented code agents.}
Repository-level agents increasingly combine code generation with external execution feedback.
Reflection provides one mechanism for converting execution feedback into subsequent actions \cite{shinn2023reflexion}.
Retrieval-augmented generation, code retrieval, and documentation retrieval offer complementary ways to ground a model in non-parametric context \cite{lewis2020retrieval,parvez2021retrieval,zhou2023docprompting}.
For example, SolAgent uses compiler and static-analysis feedback in nested refinement loops and exposes file-system tools for dependency resolution \cite{chen2026solagent}.
\benchmarkname{} does not treat such feedback loops as interchangeable model capabilities; instead, it fixes one agent scaffold and evaluates every returned patch with the same external validator and resource limits.

\paragraph{Why Pony is different.}
Pony's reference capabilities describe aliasing rights that permit efficient actor communication while statically preventing data races \cite{clebsch2015deny}.
The ORCA collector exploits these language guarantees for actor-local, concurrent memory management \cite{clebsch2017orca}.
Repository-level defects therefore include language-specific interactions among capabilities, compiler phases, actor behavior, garbage collection, and native dependencies.
Historical reproduction must also account for compiler and library evolution.
A benchmark that silently substitutes a current toolchain for an old one risks changing both compilability and semantics.

\section{Task Definition}

An instance $i$ is a tuple
\begin{equation}
  i=(R_i,c_i,D_i,P_i^{*},T_i,E_i),
\end{equation}
where $R_i$ is a repository, $c_i$ is the base commit, $D_i$ is the issue description, $P_i^{*}$ is the developer gold patch, $T_i$ is a black-box test patch, and $E_i$ is the runtime specification.
An evaluated agent receives $(R_i,c_i,D_i)$ and produces a model patch $\hat{P}_i$; it does not receive $P_i^{*}$ or the content of $T_i$.

\paragraph{Gold admissibility.}
The dataset admits an instance only when the frozen audit records
\begin{align}
  \mathrm{Exec}(R_i@c_i + T_i,E_i) &\neq 0, \\
  \mathrm{Exec}(R_i@c_i + P_i^{*} + T_i,E_i) &= 0.
\end{align}
This criterion binds the test to a behavior changed by the developer patch.
Build and infrastructure failures are not sufficient evidence of defect reproduction; construction artifacts retain the command, state-specific result, and provenance used by the audit.

\paragraph{Model resolution.}
A model resolves an instance only if $\hat{P}_i$ is nonempty, applies to the base checkout, the patched project compiles under $E_i$, and the frozen test passes:
\begin{equation}
  \begin{aligned}
  \mathrm{Resolved}(i,\hat{P}_i)
  ={}& A_i \land C_i \\
     &{}\land[\mathrm{Exec}(R_i@c_i+\hat{P}_i+T_i,E_i)=0].
  \end{aligned}
\end{equation}
Provider rejection, quota exhaustion, node failure, missing runtime assets, and validator exceptions are reported separately.
They are not converted into semantic model failures.

\section{Benchmark Construction}

\subsection{Mining and Canonicalization}

Candidate discovery combines cloned repository histories with cached GitHub issue and pull-request metadata.
For a merged pull request, the pipeline recovers the first parent of the merge commit as the base revision and the merged revision as the developer-fix state.
It retains changes that touch production Pony source outside test, documentation, example, and benchmark directories.
Pull requests whose primary intent is documentation, dependency maintenance, formatting, release engineering, or CI are removed at this stage.

The remaining pull requests are joined to issues through explicit closing relations, associated-issue metadata, issue URLs, and local issue references in pull-request or commit text.
The relation type is preserved because a closing relation provides stronger provenance than an incidental mention.
Referenced GitHub objects that resolve to another pull request, a discussion, or a missing page are not treated as issues.
Defect terms and bug labels guide candidate discovery, but they do not determine final admission: every retained instance must later satisfy the executable two-state gate.

The expanded discovery indexed 8,882 pull-request records and 775 issue records and recovered 344 joined repair pull requests.
These were merged with previously archived discovery waves and deduplicated into a 507-candidate pool.
For each candidate, the materializer stores the issue title and body, issue and pull-request URLs, base and fix commits, the complete developer diff, changed production files, candidate test files, commit date, and provenance.
The canonical duplicate key is the lowercased repository identity plus pull-request number; instance identifiers are audited independently.

\begin{table}[t]
\centering
\small
\caption{Construction lineage. Discovery and release checkpoints are shown separately because later releases union independently validated waves rather than applying one linear filter to all 507 candidates.}
\label{tab:lineage}
\begin{tabular}{@{}lrl@{}}
\toprule
Checkpoint & Count & Evidence or operation \\
\midrule
PR metadata index & 8,882 & Repository/GitHub history \\
Issue metadata index & 775 & Cached GitHub issues \\
Joined repair PRs & 344 & Explicit issue relations \\
Unique candidate pool & 507 & Cross-wave canonicalization \\
Machine release v3 & 130 & Recorded two-state passes \\
Combined release v4 & 227 & $130+110-13$ overlaps \\
Strict release v5 & 287 & v4 plus 60 new passes \\
Frozen release v6 & 291 & v5 plus four new passes \\
\bottomrule
\end{tabular}
\end{table}

\subsection{Executable Acceptance and Release Lineage}

Candidate materialization and executable admission are separate.
Materialization establishes a reproducible task identity and obtains a nonempty developer patch and test patch.
Admission then checks the same test against two repository states: the base state must return a nonzero exit code, while the base state with the developer patch must return zero.
The command, runtime mapping, exit codes, logs, and patch identities are retained so that a row is not accepted merely because its issue and pull request look semantically related.

The final size of 291 arises from versioned unions of independently validated waves, not from choosing 291 as a target.
The v3 machine release contains 130 rows.
The v4 union combines those rows with a 110-row strict two-state source and removes 13 overlapping identities, yielding 227 unique instances.
The v5 expansion contributes 60 non-overlapping passes from a 73-row validated source; the other 13 are canonical repository--PR duplicates, producing 287 instances.
Finally, four newly accepted and non-overlapping rows form v6, so $287+4=291$.

The v6 release audit is an offline merge and schema-normalization audit rather than another test execution.
It verifies the recorded two-state evidence for all 291 source rows, checks that every required field and both patches are nonempty, and reports 291 unique canonical repository--PR pairs and 291 unique instance identifiers.
Its output is hash-locked, allowing later runtime enrichment and model evaluation to refer to an immutable task set.

\begin{algorithm}[t]
\caption{\benchmarkname{} construction and release}
\label{alg:construction}
\begin{algorithmic}[1]
\REQUIRE Merged-PR metadata $M$, issue metadata $I$, archived validated waves $A$
\ENSURE Frozen executable dataset $D$
\STATE $C \leftarrow \emptyset$
\FOR{each merged pull request $p \in M$}
    \STATE Recover base and fix revisions from repository history
    \IF{$p$ changes production Pony code and links a valid issue}
        \STATE Materialize issue text, patches, files, revisions, and provenance
        \STATE Add the canonical candidate to $C$
    \ENDIF
\ENDFOR
\FOR{each candidate $c \in C$}
    \STATE Acquire an official test or generate and review a black-box test
    \STATE Bind $c$ to its historical runtime and validation command
    \IF{base fails and developer fix passes the same test}
        \STATE Add $c$ and its execution evidence to validated wave $V$
    \ENDIF
\ENDFOR
\STATE $D \leftarrow$ canonical non-overlapping union of $A$ and $V$
\STATE Audit schema, evidence, identities, counts, and SHA-256
\RETURN $D$
\end{algorithmic}
\end{algorithm}

\subsection{Test Acquisition and Review}

The construction process prefers test behavior already present in the fixing pull request.
Of 291 instances, 185 use tests recorded as \texttt{original\_pr}, 97 use an official PR test change normalized for the benchmark, and six use another official PR test route.
Only three instances use generated tests.
Thus 288 instances (98.97\%) trace their oracle to developer or official PR test evidence, while generated tests account for 1.03\%.

For an official-test instance, the construction pipeline separates production changes from the test change so that the latter can be applied independently to the historical base state.
The resulting \texttt{test\_patch} is kept hidden during model interaction.
Where an official test depends on unrelated pull-request structure, the benchmark records a normalized official route rather than silently mixing the developer fix into the oracle.
Each row retains its test origin, source pull request, test files, execution command, and validation record; the three origin categories are therefore auditable rather than inferred from patch content.

Generated tests follow a writer--review--execute loop adapted from ArkEval \cite{xie2026arkeval}.
The writer receives the issue, base revision, gold diff, repository context, and a versioned test-writing specification.
Independent reviewers check whether the test exercises public behavior, avoids reading the implementation or gold patch, and distinguishes the reported defect from unrelated compilation or environment errors.
Any requested rewrite invalidates the round; the revised test must pass the complete gate again.
The final release records test origin rather than presenting generated and developer tests as interchangeable evidence.

\subsection{Two-State Execution Gate}

Every oracle is executed under the runtime mapped to the candidate's base commit.
The validator first checks out a clean base state, applies only the test patch, and runs the recorded command.
It then restores the same base, applies the developer patch followed by the identical test patch, and reruns that command under the same runtime.
An instance is admitted only when the first state fails and the second passes.
Patch-application errors, missing dependencies, unavailable images, compiler setup failures, and validator exceptions are recorded as construction failures rather than evidence that the issue was reproduced.

This gate serves two purposes.
First, it rejects tests that merely exercise already-correct behavior because such a test would pass on the base revision.
Second, it rejects tests that fail for reasons the developer patch does not repair because the fixed state would remain nonzero.
The gate does not prove that an oracle completely specifies the issue, which motivates the independent semantic review described later, but it provides a uniform executable lower bound for admission.

The release audit preserves the state-specific exits instead of collapsing them into a single ``validated'' flag.
Across the 291 rows, historical evidence appears under several versioned schema paths because the dataset was built in waves.
The v6 audit normalizes these representations, verifies that every row yields the same base-fail/fix-pass predicate, and reports 291 passes and zero failures.
Algorithm~\ref{alg:construction} summarizes the complete path from GitHub history to the frozen release.

\subsection{Historical Runtime Reconstruction}

Pony repositories span more than a decade of compiler and dependency changes.
We therefore map every instance to an explicit runtime definition instead of using one contemporary compiler.
The evaluation input contains 291 runtime mappings and 72 unique historical images.
The images cover 289 unique base commits; the two fewer unique commits arise because multiple issues can share the same repository state.

Runtime reconstruction is checked independently on five nodes: one x86\_64 V100 server, one aarch64 Huawei server, and three aarch64 DGX Spark systems.
Every node passes a 72/72 image handshake.
The Huawei path uses PRoot, QEMU user emulation, and a glibc 2.27 compatibility layer for images that require glibc 2.23.
These adaptations are part of the frozen environment manifest and must be reproduced rather than hidden behind a generic ``Docker available'' statement.

\subsection{Dataset Characteristics}

\begin{table}[t]
\centering
\small
\caption{\benchmarkname{} v6 characteristics. Patch sizes count diff lines.}
\label{tab:dataset}
\begin{tabular}{@{}lr@{}}
\toprule
Characteristic & Value \\
\midrule
Instances & 291 \\
Repositories & 15 \\
Unique base commits & 289 \\
Historical runtime images & 72 \\
Commit date range & 2015--2026 \\
Median gold-patch lines & 92 \\
Mean gold-patch lines & 250.25 \\
Median test-patch lines & 93 \\
Mean test-patch lines & 196.42 \\
Median defect files & 1 \\
Developer/official test origins & 288 (98.97\%) \\
Generated test origins & 3 (1.03\%) \\
\bottomrule
\end{tabular}
\end{table}

\begin{table}[t]
\centering
\small
\caption{Repository distribution. The concentration in \texttt{ponyc} is reported as a limitation rather than hidden by aggregation.}
\label{tab:repos}
\begin{tabular}{@{}lr@{}}
\toprule
Repository & Instances \\
\midrule
\texttt{ponylang/ponyc} & 211 \\
\texttt{ponylang/lori} & 17 \\
\texttt{ponylang/postgres} & 17 \\
\texttt{ponylang/ssl} & 13 \\
\texttt{ponylang/stallion} & 12 \\
\texttt{ponylang/hobby} & 8 \\
Other nine repositories & 13 \\
\bottomrule
\end{tabular}
\end{table}

The benchmark is deep rather than balanced: \texttt{ponylang/ponyc} contributes 211 instances (72.51\%).
The median gold patch touches one defect file, but the maximum is 44, and the longest gold diff contains 7,583 lines.
These tails preserve repository-level tasks that cannot be reduced to a single local edit.

\begin{table*}[!t]
\centering
\scriptsize
\caption{Five-model results. Every percentage uses that model's actual nonempty Patch count as the denominator; no values are extrapolated to 291 patches. Repro and Resolved coincide under the frozen issue-specific oracle.}
\label{tab:results}
\begin{tabular}{@{}lrrrrr@{}}
\toprule
Model & Localization & Apply & Build & Repro & Resolved \\
\midrule
GPT-5.6-sol & 96.91\% & 100.00\% & 64.95\% & 18.21\% & 18.21\% \\
DeepSeek-V4-Pro & 91.91\% & 99.57\% & 54.47\% & 10.21\% & 10.21\% \\
GLM-5.2 & 93.95\% & 100.00\% & 49.19\% & 18.55\% & 18.55\% \\
MiniMax-M3 & 96.00\% & 100.00\% & 38.40\% & 12.00\% & 12.00\% \\
Kimi-K3 & 97.40\% & 100.00\% & 79.22\% & 24.68\% & 24.68\% \\
\bottomrule
\end{tabular}
\end{table*}

\section{Evaluation Protocol}

\subsection{Research Questions}

We organize the evaluation around three questions:
\begin{itemize}
    \item \textbf{RQ1:} What repository, temporal, patch, and runtime diversity does \benchmarkname{} provide?
    \item \textbf{RQ2:} Under one matched agent scaffold, how often do current models produce applicable, compilable, test-passing Pony patches?
    \item \textbf{RQ3:} Which stages---generation, application, compilation, semantic testing, provider access, or infrastructure---account for observed failures?
\end{itemize}

\subsection{Agent and Models}

All models run through mini-SWE-agent 2.4.6 \cite{minisweagent246}.
Every model receives the same runtime-enhanced 291-instance input and benchmark configuration.

The compared models are GPT-5.6-sol, DeepSeek-V4-Pro, GLM-5.2, MiniMax-M3, and Kimi-K3.
Each model uses ten workers, a 120-step limit, and a 3,600-second wall-time limit per instance.
API credentials are process-local and are excluded from repository files and logs.
Patch generation does not use the gold patch or hidden test patch.

\subsection{Strict Validation and Metrics}

The primary quality metric is \textbf{Resolved/Patch}, the fraction of a model's actual nonempty patches that pass the strict model-resolution predicate.
We additionally report:
\begin{enumerate}
    \item localization rate;
    \item patch-application rate;
    \item compilation rate;
    \item hidden-test pass rate;
    \item provider-failure and infrastructure-failure counts;
    \item median wall time for terminal, non-infrastructure cases.
\end{enumerate}

For localization, application, compilation, reproduction, and resolution, the denominator is the actual Patch count for that model.
These conditional rates are not extrapolated into hypothetical results over 291 patches.
Provider and infrastructure failures remain generation outcomes rather than semantic patch failures.

\section{Results}

\subsection{RQ1: Benchmark Composition}

\benchmarkname{} contains 291 tasks that pass the frozen executable audit across 15 repositories and 289 unique historical base commits.
The date range extends from September 2015 to July 2026.
Most tasks are anchored by developer or official PR tests, but the benchmark remains concentrated in the compiler repository.
The runtime layer is comparably diverse: 72 images are required to replay the full set, and all five nodes pass the corresponding environment handshake.
These findings support the benchmark's historical executability while delimiting its ecosystem coverage.

\subsection{RQ2: Five-Model Repair Effectiveness}

Table~\ref{tab:results} reports patch-conditional percentages without extrapolating ungenerated patches.
Kimi-K3 attains the highest build rate (79.22\%) and resolution rate (24.68\%).
GLM-5.2 and GPT-5.6-sol follow with resolution rates of 18.55\% and 18.21\%, while MiniMax-M3 and DeepSeek-V4-Pro reach 12.00\% and 10.21\%, respectively.
These comparisons concern only the patches actually produced by each model and should not be interpreted as full-dataset resolution rates.

\subsection{RQ3: Failure Attribution}

Stage-separated accounting reveals substantially different bottlenecks.
Patch application is nearly saturated for every model, ranging from 99.57\% to 100.00\%, whereas compilation ranges from 38.40\% to 79.22\%.
Even after compilation, many patches fail the issue-specific oracle: the final resolution rates range from 10.21\% to 24.68\%.
Thus syntactic applicability is not a useful proxy for executable repair, and historical compilation and behavioral validation remain the dominant filters.

\section{Independent Full-Set Selection Review}

We measure agreement on whether each benchmark instance should be selected, not on manuscript prose.
Three isolated machine reviewers independently reassessed all 291 frozen instances in JSONL order without treating release eligibility fields as ground truth.
Under a shared binary rubric, \emph{include} requires a source-code task with an observable outcome---either a defect repair or a bounded feature request---alignment between the issue and developer fix, and a test of public or user-observable behavior.
\emph{Exclude} covers documentation, CI, dependency-only, formatting, or behavior-preserving refactoring tasks; issue--gold semantic mismatches; tests coupled only to private implementation details; and insufficient evidence.
The frozen v6 audit supplies the execution gate, so stale per-row status fields or historical dependency logs cannot by themselves trigger exclusion.
Before full annotation, the reviewers completed a 30-instance high-ambiguity calibration round ($\kappa=0.8582$); the resulting guideline fixed the treatment of pure API renames, private-method-only tests, and alternatives explicitly permitted by an issue.

This review is deliberately downstream of release construction.
The executable gate answers whether a frozen test distinguishes the base and developer-fix states; the selection review asks whether the issue, fix, and test jointly represent a suitable public-behavior software task.
Consequently, the review diagnoses semantic composition but does not produce the number 291 or silently delete rows from the frozen release.
Isolation prevents one reviewer's labels from anchoring another, while the shared calibration makes the decision boundary explicit before the full pass.

\begin{table}[t]
\centering
\small
\caption{Independent machine review of all 291 instances.}
\label{tab:selection-review}
\begin{tabular}{@{}lrr@{}}
\toprule
Reviewer & Include & Exclude \\
\midrule
A & 253 & 38 \\
B & 255 & 36 \\
C & 254 & 37 \\
\midrule
All 873 ratings & 762 & 111 \\
\bottomrule
\end{tabular}
\end{table}

The reviewers unanimously include 249 instances and unanimously exclude 32, yielding unanimous item-level agreement on 281/291 instances (96.56\%).
The remaining ten instances split two-to-one: five toward inclusion and five toward exclusion.
Majority labels are therefore include for 254 instances and exclude for 37.
We use Fleiss' $\kappa$ because three reviewers assign the same binary categories.
For item $i$, let $n_{iY}$ and $n_{iN}$ be the numbers of include and exclude labels.
Its observed agreement is
\begin{equation}
 P_i=\frac{n_{iY}(n_{iY}-1)+n_{iN}(n_{iN}-1)}{3(3-1)}.
\end{equation}
Across all items, $\bar{P}=0.9771$.
The 873 labels contain 762 inclusions and 111 exclusions, giving marginal chance agreement
\begin{equation}
 P_e=(762/873)^2+(111/873)^2=0.7780.
\end{equation}
Thus Fleiss' $\kappa=(\bar{P}-P_e)/(1-P_e)=0.8968$.

This is a machine-review diagnostic, not human-expert agreement or final dataset adjudication.
We therefore keep the frozen artifact immutable and route the 32 unanimous exclusions and ten split decisions to human and execution-evidence adjudication.
The review complements rather than replaces the executable two-state gate.

\section{Discussion}

\paragraph{Executable and semantic validity.}
\benchmarkname{} separates executable validity from semantic validity.
The two-state gate asks whether a test distinguishes the historical base from the developer fix; the selection review asks whether the issue, production change, and test form an appropriate public-behavior repair task.
A high $\kappa$ shows consistent application of the machine-review rubric, not label correctness or final human approval.
Conversely, semantic disagreement does not erase recorded execution evidence.
Reporting both layers avoids hiding either judgment inside one release flag.

\paragraph{Interpreting model percentages.}
Using actual nonempty patches as the denominator measures patch quality conditional on generation and exposes stage-to-stage degradation without inventing missing outcomes.
These percentages are not end-to-end success rates over 291 tasks and must not be used to claim total provider coverage.
Generation, provider, and infrastructure outcomes therefore remain separate.

\paragraph{Stable, reproducible curation.}
The frozen v6 file remains immutable so predictions and validation records retain stable identities.
After human adjudication, a curated subset should be a new version with an explicit v6 mapping, SHA-256, and inclusion log, enabling sensitivity analysis without retroactive deletion.
Reproduction also binds each row to its repository, base commit, developer and test patches, runtime image, and command; model runs additionally bind the agent commit, configuration, limits, and prediction artifact.
Manifests and hashes are therefore part of the protocol.

\section{Threats to Validity}

\paragraph{Repository concentration.}
The compiler repository contributes 72.51\% of instances.
Results may therefore characterize repair of Pony's compiler and standard-library ecosystem more strongly than arbitrary Pony applications.
We report per-repository outcomes where denominators permit, but do not claim balanced ecosystem coverage.

\paragraph{Oracle validity.}
Fail-to-pass behavior establishes that a test distinguishes the base and developer-fix states; it does not prove that the test fully specifies the issue.
Developer tests can be narrow, and generated tests can overfit visible patch details.
PonyEval mitigates this risk with black-box constraints, explicit test provenance, full-set independent machine review, stricter review for generated tests, and preservation of complete execution evidence.
High machine agreement does not establish correctness, and the 32 unanimous exclusions plus ten split decisions show that executable two-state evidence alone does not settle semantic admissibility.
The three generated tests should also be analyzed separately in sensitivity checks.

\paragraph{Historical environment fidelity.}
Pinned images improve reproducibility but can still differ from the original developer machine, especially when emulation or compatibility layers are required.
The five-node handshake tests runtime availability, not bitwise equivalence of every system component.

\paragraph{Model and provider instability.}
The evaluated model names identify provider endpoints at a particular time; model weights and serving behavior may change without public version hashes.
We therefore preserve request-independent configuration, predictions, trajectories, timestamps, and output hashes.
Quota and authentication failures are reported as provider failures.

\paragraph{Contamination and temporal effects.}
The benchmark uses public GitHub history through July 2026, so the degree to which proprietary models have seen individual issues is unknown.
PonyEval measures end-to-end repair under a fixed agent protocol, not uncontaminated synthesis ability.
Future releases should add a post-training-date or live split where model cutoffs are documented.

\section{Conclusion}

We introduced \benchmarkname{}, a 291-instance, executable benchmark for real-world Pony issue resolution.
The benchmark binds natural-language issues to historical repository states, developer fixes, black-box tests, and explicit runtime mappings.
Its frozen audit reports complete two-state gold validation and zero duplicate identities; its execution layer reconstructs 72 runtime images and verifies them on five heterogeneous nodes.
The independent 291-instance binary selection review yields Fleiss' $\kappa=0.8968$; its 32 unanimous exclusions and ten split decisions remain explicit adjudication items rather than silent dataset revisions.
The matched mini-SWE-agent evaluation compares five current models using each model's actual generated patches as the denominator.
Kimi-K3 records the highest conditional resolution rate (24.68\%), followed by GLM-5.2 (18.55\%) and GPT-5.6-sol (18.21\%).
The large drop from application to compilation and behavioral success confirms that executable validation is essential for evaluating Pony repair.

\paragraph{Data and artifact availability.}
The public release artifact will include the frozen dataset, audit manifests, test patches, runtime mappings, mini-SWE-agent configuration, validator, prediction files, and SHA-256 index.
The archival URL will be added when the artifact is deposited.

\clearpage
\bibliography{ponyeval_references}

\end{document}